%% file: main.tex
\documentclass{edm_article}

\usepackage{graphicx}
\usepackage[acronym]{glossaries}
\usepackage{subfig}
\usepackage{multirow}

\newcommand{\spara}[1]{\smallskip\noindent{\bf{#1}}}

\newacronym{SUA}{SUA}{Sistema \'{U}nico de Admisi\'{o}n}
\newacronym{LDA}{LDA}{Latent Dirichlet Allocation}
\newacronym{CRUCH}{CRUCH}{Consejo de Rectores de las Universidades Chilenas}
\newacronym{PSU}{PSU}{Prueba de Selecci\'{o}n Universitaria}
\newacronym{DEMRE}{DEMRE}{Departamento de Evaluaci\'{o}n, Medici\'{o}n y Registro Educacional}
\newacronym{NEM}{NEM}{Notas de Ense\~{n}anza Media}
\newacronym{OECD}{OECD}{Organisation for Economic Co-operation and Development}
\newacronym{UN}{UN}{United Nations}
\newacronym{STEM}{STEM}{Science, Technology, Engineering and Mathematics}
\newacronym{SAT}{SAT}{Scholastic Assessment Test}
\newacronym{SES}{SES}{socioeconomic status}
\newacronym{DA}{DA}{Deferred Acceptance}
\newacronym{TTC}{TTC}{Top Trading Cycles}
\newacronym{CSV}{CSV}{comma-separated values}
\newacronym{WYSIWYG}{WYSIWYG}{{\it what you see is what you get}}
\newacronym{WAE}{WAE}{{\it we are all equal}}
\newacronym{EEOC}{EEOC}{Equal Employment Opportunity Commission}
\newacronym{AUC}{AUC}{Area under the Curve}
\newacronym{ROC}{ROC}{receiver operating characteristic}
\newacronym{DCG}{DCG}{discounted cumulative gain}
\newacronym{nDCG}{nDCG}{normalized discounted cumulative gain}
\newacronym{NLP}{NLP}{natural language processing}
\newacronym{GP}{GP}{Gaussian process}
\newacronym{TPE}{TPE}{Tree-structured Parzen Estimator}
\newacronym{SMBO}{SMBO}{Sequential Model-Based Global Optimization}
\newacronym{SPD}{SPD}{Statistical Parity Difference}
\newacronym{DI}{DI}{disparate impact}
\newacronym{GPA}{GPA}{Grad Point Average}
\newacronym{EBP}{EBP}{evidence-based policy}
\newacronym{MAE}{MAE}{mean absolute error}
\newacronym{MLPC}{MLPC}{Multi-Label Probabilistic Classifier}
\newacronym{LTR}{LTR}{Learning-to-Rank}
\newacronym{MOR}{MOR}{Multi-Output Regression}
\newacronym{SD}{SD}{standard distribution}

\begin{document}

\title{Towards Data-Driven Affirmative Action Policies under~Uncertainty}
%
\numberofauthors{1}
\author{
Corinna Hertweck,$^1$\thanks{Work done while at the University of Helsinki.}\enspace Carlos Castillo,$^2$ Michael Mathioudakis$^3$ 
\and
\affaddr{$^1$ Zurich University of Applied Sciences, $^2$ Universitat Pompeu Fabra, $^3$ University of Helsinki}\\
}

\maketitle


\begin{abstract}
In this paper, we study university admissions under a centralized system that uses grades and standardized test scores to match applicants to university programs.
We consider affirmative action policies that seek to increase the number of admitted applicants from underrepresented groups. 
Since such a policy has to be announced before the start of the application period, there is uncertainty about the score distribution of the students applying to each program.
This poses a difficult challenge for policy-makers.
We explore the possibility of using a predictive model trained on historical data to help optimize the parameters of such policies.
\end{abstract}

%


\input{chapters/01_introduction}

\input{chapters/01a_related_work}
\input{chapters/02_data}
\input{chapters/03_policy_models}
\input{chapters/04_evaluation}
\input{chapters/05_conclusions}


%

\small

\bibliographystyle{abbrv}
\bibliography{main}  
%
%
\balancecolumns
\end{document}

%% file: chapters/01_introduction.tex
\section{Introduction}\label{sec:introduction}

In centralized university admission systems, students submit their applications to a central institution, which matches the applicants with the programs offered by various universities.
This matching is typically automated through an algorithm and based on grades and standardized test scores.
As grades and test scores tend to differ across demographic groups (see, e.g.,~\cite{bacharach2003racial, mcewan2004indigenous, reardon2013widening, rothstein2015racial}), this kind of system can lead to large gaps in admission rates between groups.
These gaps can be reduced through affirmative action policies, which are challenging to design.
First, such policies should reward merit and lead to the admission of the best applicants.
Second, policies should be announced \emph{before} the application period begins to provide candidates with all the information they need.
Computational methods can be leveraged to evaluate a large range of alternatives, much more than can be evaluated manually, and identify effective policies.

In this paper, we develop an approach for designing robust and effective affirmative action policies for university admissions. 
We consider {\it bonus policies}, i.e., policies that add a number of bonus points to the scores of applicants from disadvantaged backgrounds.  
These policies do not alter the admission priority of applicants within each group, and have equivalent effects to setting admission quotas~\cite{mathioudakis2019affirmative}.
The technical problem we face is to choose the right number of bonus points so that the policy will have a consistently beneficial effect on the admission rate of the given group.

%% file: chapters/01a_related_work.tex
\spara{Related Work}

Using the terminology of Friedler et~al.~\cite{friedler2016possibility}, our work and affirmative action more generally are consistent with the \gls{WAE} worldview.
According to \gls{WAE}, higher grades do not necessarily imply more talent or more diligence, but are seen as a product of structural inequality (e.g., lack of resources for subgroups of the student population~\cite{evans2004environment}).
Affirmative action policies (e.g., bonus policies) are thus implemented as a way to counter the effect of inequalities being reinforced through -- in our case -- unequal grade and test score distributions.
The beneficial effects of affirmative action policies on disadvantaged groups have both been shown in studies based on real-world implementations of specific policies (e.g., \cite{Bastarrica2018, feedback-california}) and in laboratory settings (e.g., \cite{women-competition}).
Other studies discuss how different types of affirmative action policies can be incorporated into the algorithm that matches students with university programs (see, e.g., \cite{abdulkadirouglu2005college, effective-school-choice, skipping-down}).
However, existing literature lacks a discussion of how the numerical parameters in these policies (e.g., the number of bonus points given to an underrepresented group) affect the resulting matching.
The usage of computational methods in policy design has, for example, already been researched in the fields of public health~\cite{barbrook2017uses} or counter-terrorism~\cite{keller2010dismantling,meltzer2001modeling}.
In the area of university admissions, recent work~\cite{mathioudakis2019affirmative} addressed the problem of the design of affirmative action policies.
However, the study designed policies for a given set of applications and not under uncertainty.
To our knowledge, this is the first study that discusses how historical data can be leveraged in the design of affirmative action policies for university admissions when the applicants are unknown.

%% file: chapters/02_data.tex
\section{Dataset}\label{sec:setting}

We analyze data from the university admission system of Chile, 
which contains information on all applicants and university programs between 2004 and 2017. 
For 2017, this amounts to 60\,000 students and about 1\,500 programs.

After taking standardized tests and obtaining their scores, students submit their \emph{preferences} as a ranked list of up to ten programs.
Students are admitted to programs according to an admission score, which is a weighted average of high school grades and test scores -- with different weights used by different programs.
The data contains these high school grades, test scores, and applicant preferences.
It also includes a (binary) gender variable as well as the household income and size.
We binarize the student's income as {\it low-income} (bottom five household-income-per-household-member deciles), and {\it high-income} (top five deciles).

\begin{figure}
    \begin{center}
    \includegraphics[width=.40\textwidth, trim=0 0 0 0, clip]{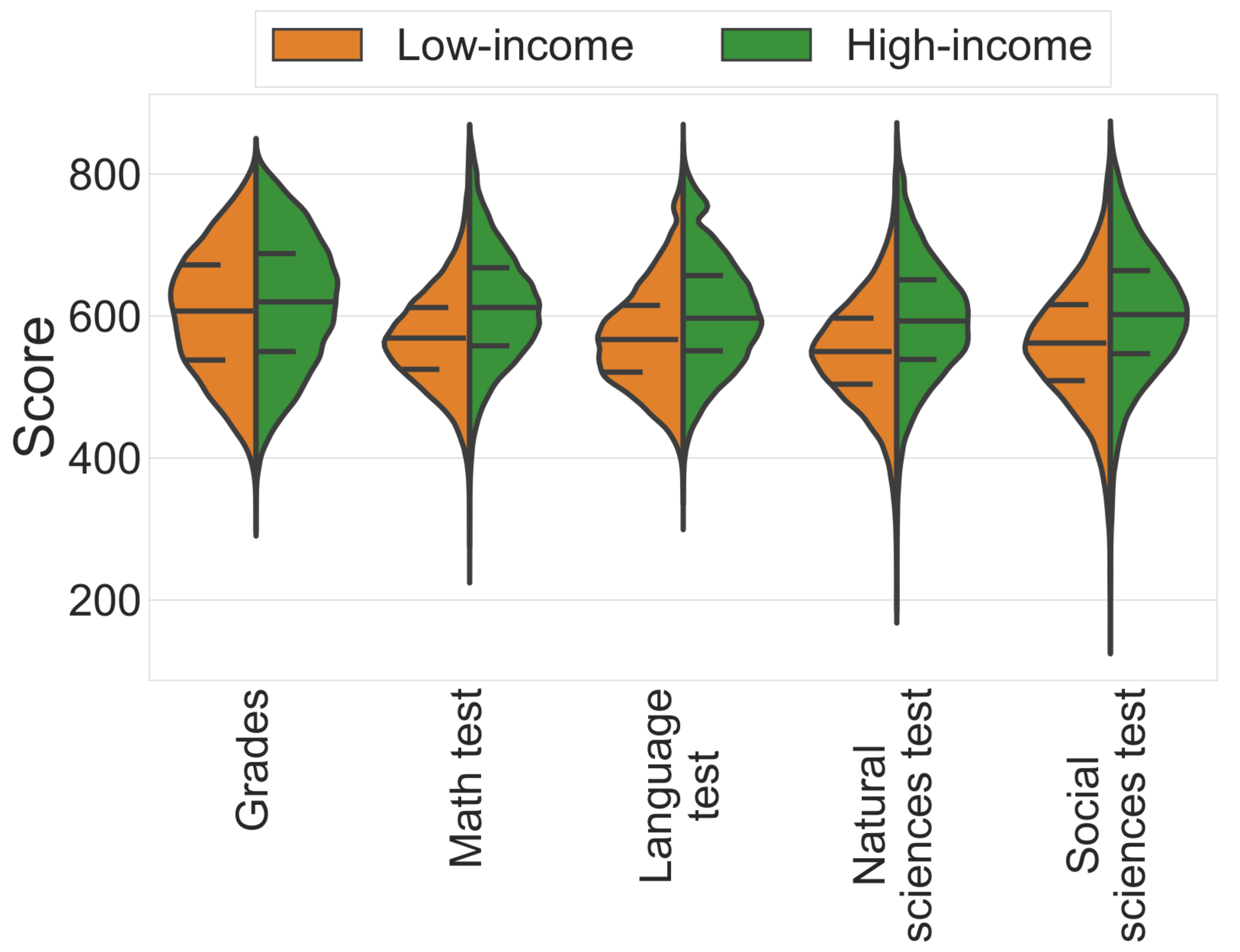} 
    \end{center}
    \caption{Distributions of grades and standardized test scores for different income levels.}
    \label{fig:compare_scores_by_income}
\end{figure}

\spara{Differences in scores and preferences.}
According to statistics of the year 2016, with similar observations for other years, 
we find that there is a pronounced difference between the score distributions of different income groups: high-income students have slightly higher high school grades than low-income students, and much higher scores across all standardized tests (Figure~\ref{fig:compare_scores_by_income}). 
While there exist differences between genders, too, they are less pronounced in the data.

\begin{figure}
    \centering
    \includegraphics[width=.40\textwidth, trim=0 0 0 0, clip]{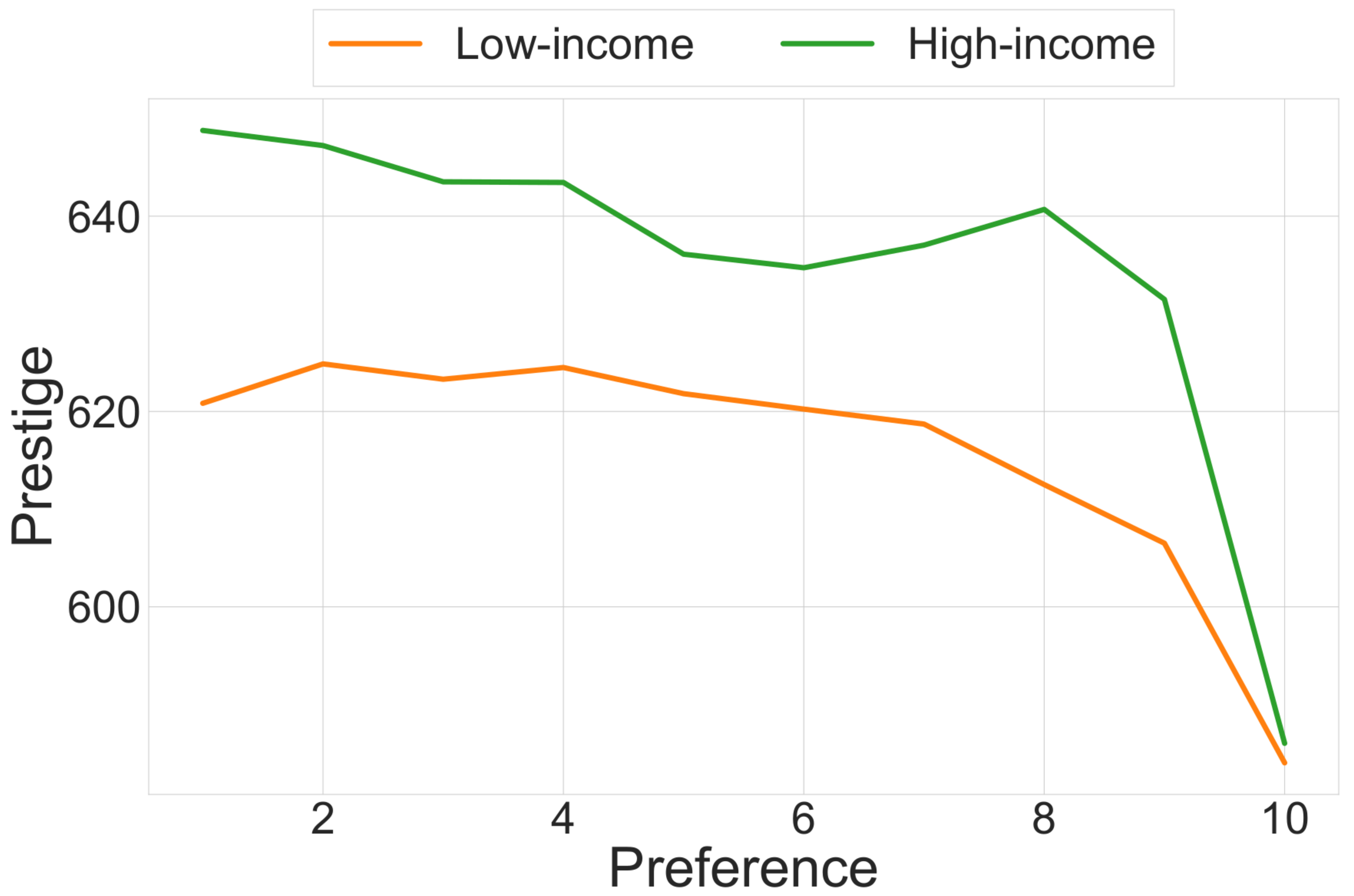}
    \caption{Average prestige of students' preferences across income levels.}
    \label{fig:prestige_by_preference_by_income}
\end{figure}

We also observe differences in the programs to which students apply.
To describe these differences, we define the \emph{prestige} of a program as the average score of its admitted students in the previous three years.
We find that high-income students express preference for programs of higher average prestige compared to low-income students, particularly in their first preference (Figure~\ref{fig:prestige_by_preference_by_income}).
The difference can largely be attributed to the different scores of the two groups (Figure~\ref{fig:compare_scores_by_income}) -- once we control for admission scores, the differences mostly disappear (figure omitted for brevity).

\begin{figure}
\centering
\includegraphics[width=.4\textwidth, trim=0 0 0 0, clip]{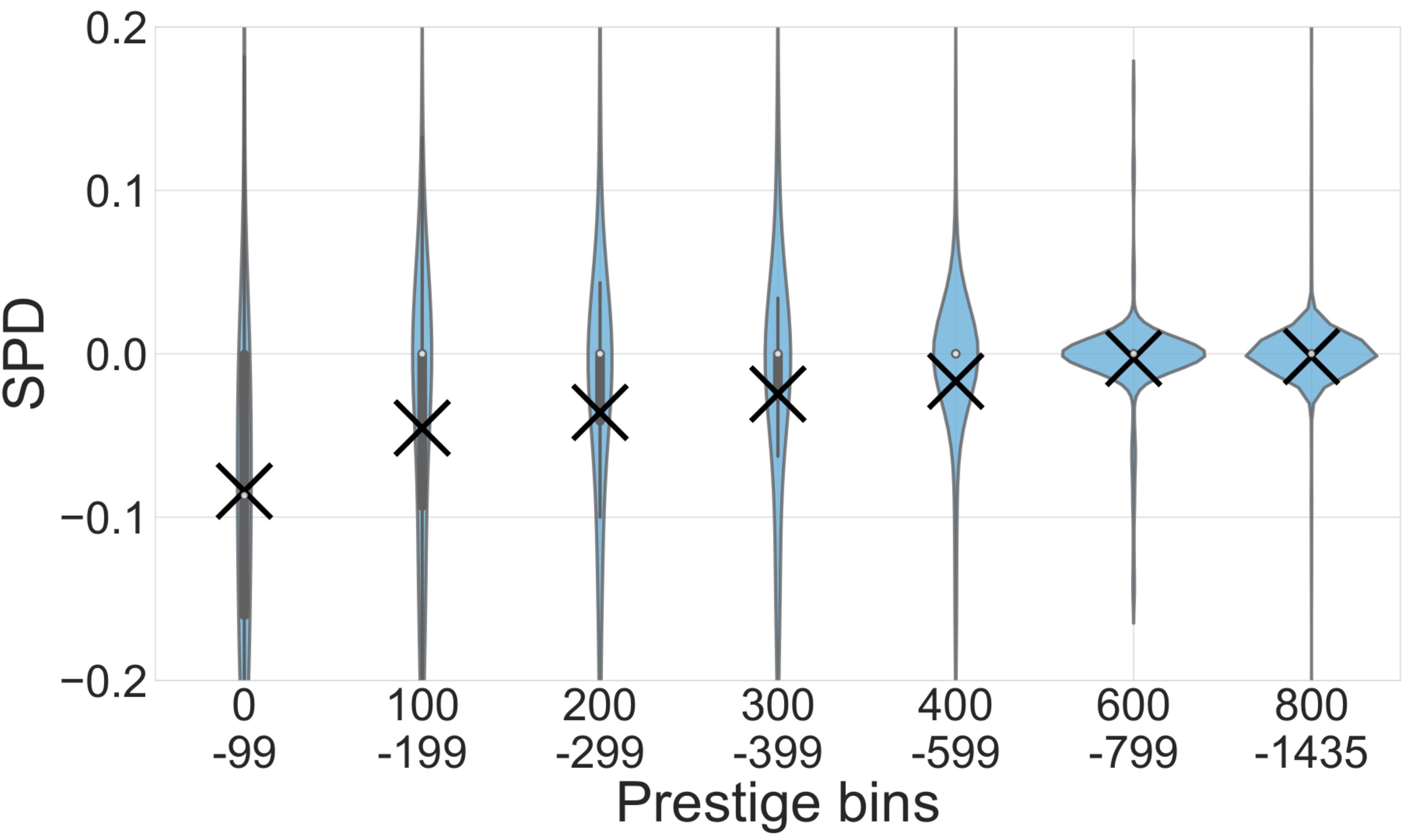}
\caption{Distribution of \gls{SPD} for income.
Programs are ranked by prestige (higher to lower); x-axis indicates program ranks in each bin; crosses mark the means.}
\label{fig:spd_income}
\end{figure}

\spara{Differences in admission rates.}
To measure differences in admission rates, we use the \textit{\gls{SPD}} measure:
\begin{equation}\label{eq:statistical-parity-difference}
    P(Y = 1 | A = a) - P(Y = 1 | A \neq a),
\end{equation}
where $Y=1$ indicates being admitted into a program, $A$ is a sensitive attribute, and
$a$ marks a demographic group (in this work either low-income or female students).
Low absolute values of \gls{SPD} are desired -- and perfect equality is achieved for a value of 0. 
Following thresholds provided in tools for measuring algorithmic bias~\cite{aif360-oct-2018}, we accept values inside $[-0.1, 0.1]$ and reject values outside of this range as \emph{strongly unequal}.
As shown in Figure~\ref{fig:spd_income}, we tend to measure larger negative \gls{SPD} values for the more prestigious programs, indicating differences in admission rates that place low-income students at a disadvantage.

\spara{Variance in admission rates across years}. 
We seek to bring the \gls{SPD} of programs closer to zero, which is challenging because
while the share of programs with strongly unequal admission rates is fairly constant across years, there is large variation in the \gls{SPD} of individual programs over the years.
In fact, we found that the \gls{SPD} of one program in one year is a worse predictor for its \gls{SPD} in the next year than simply predicting an \gls{SPD} of 0.
This observation is important in practice, as we could be under-correcting or over-correcting inequalities if we assume differences in admission rates will not vary.

%% file: chapters/03_policy_models.tex
\section{Policy Design}\label{sec:policy-models}

To reduce admission rate gaps, we consider the use of bonus policies, which award bonus points to students from disadvantaged groups.

\spara{Problem definition}. 
We wish our admission policy to lead to the admission of the students with the highest scores, and to a reduction in admission rate disparities.
We thus define the objective function of the bonus policy as a linear combination of both the equality of admission rates, i.e., the \gls{SPD}, and \emph{utility}, i.e., the scores of the accepted students.
We measure utility $\mu_b$ as the average score of the students who are admitted when $b$ bonus points are given.
As utility might vary strongly between application sets, we calculate the loss of $\mu_b$ compared to $\mu_0$, i.e. the utility when no bonus policy is implemented.
\begin{equation}\label{eq:objective-single}
    o_b = (\mu_0 - \mu_b) + \lambda \cdot |\text{SPD}_b|, \quad \lambda \geq 0.
\end{equation}

The optimization is executed by finding the optimal bonus policy for multiple \emph{application sets}, which are forecasts of potential applicants and their preferences based on historical data.
The optimal bonus for a single application set is determined by applying a range of bonus points and directly measuring the objective function.
The output of this procedure is simply the average of the optimal bonuses over all application sets that are evaluated.
The strategies we consider next either sample application sets from a statistical model learned from historical data; or optimize directly on historical data.

\spara{Application sets generated by a predictive model.}
We build a multi-label probabilistic classifier to model students' application behavior.
To predict applications for 2017, we train the model on data from the at the time most recent year, i.e., 2016.
We deploy the trained model to create $n$ possible application sets.
Each application set consists of a sampled set of students together with their predicted program preferences.
We experiment with $n=50$ and $n=200$ sampled application sets to evaluate possible bonus policies.

\spara{Baseline: directly use historical data.}\label{sec:years}
A simpler approach is to compute what bonus policy would have been optimal in previous years.
Such an approach has the disadvantage that we are limited to the number of years for which we have data for the program that we want to design a policy for. 
Nevertheless, we also include this approach in our empirical evaluation as a baseline.

%% file: chapters/04_evaluation.tex
\section{Experimental Evaluation}\label{sec:strategies-comparison}

We compare strategies based on application sets generated by a predictive model (50 and 200 application sets) and strategies that average the (in hindsight) optimal policies for historical data (1 year, 3 years, and 5 years). 
We consider two sensitive attributes, gender and income, for the year 2017.
To find an appropriate value for $\lambda$ in Eq.~\ref{eq:objective-single}, we calculate the median differences in grades and test scores between subgroups: $23$ points for gender groups and $28$ points for income groups.
First, we evaluate the strategies for suggesting policies for all programs and evaluate their overall effect.
Second, we evaluate the strategies when policies are only applied to programs which show consistent inequalities in admission rates over time.

\spara{All programs}.
In Table~\ref{tab:baselines_differences_objective_function}, we compare the objective function values resulting from applying the different strategies to the ideal objective function values for all programs. In the table itself, we show the mean and \gls{SD} of this difference for both sensitive attributes.
Table~\ref{tab:baselines_differences_objective_function} shows the error in the objective function relative to the smallest achievable ones, i.e. the values that result from applying the ideal bonus policies.
Note that, based on Table~\ref{tab:baselines_differences_objective_function}, policies that use more application sets for their suggestions generally lead to both smaller errors and less variance in the error.

\begin{table*}[h]
\scriptsize
\centering
\caption{Error in objective function relative to ideal policies. Smaller values are better.}
\label{tab:baselines_differences_objective_function}
\begin{tabular}{l|l|l|l|l|l|l|l|l|}
\cline{2-9}
                                            & \multicolumn{4}{l|}{All programs}                                                                                                 & \multicolumn{4}{l|}{Consistently unequal}                                                                                         \\ \cline{2-9} 
                                            & \multicolumn{2}{l|}{Gender}                                     & \multicolumn{2}{l|}{Income}                                     & \multicolumn{2}{l|}{Gender}                                     & \multicolumn{2}{l|}{Income}                                     \\ \hline
\multicolumn{1}{|l|}{Strategy}              & Mean                           & SD                             & Mean                           & SD                             & Mean                           & SD                             & Mean                           & SD                             \\ \hline
\multicolumn{1}{|l|}{Historical - 1 year}   & 0.37                           & 1.11                           & 0.44                           & 1.31                           & 1.31                           & 1.88                           & 1.79                           & 1.63                           \\ \hline
\multicolumn{1}{|l|}{Historical - 3 years}  & 0.30                           & 0.94                           & 0.34                           & 1.06                           & 1.02                           & 1.81                           & \textbf{1.36} & 1.62                           \\ \hline
\multicolumn{1}{|l|}{Historical - 5 years}  & 0.32                           & 0.99                           & \textbf{0.33} & \textbf{0.98} & 1.22                           & 1.20                           & 1.63                           & \textbf{1.57} \\ \hline
\multicolumn{1}{|l|}{Predictive - 50 sets}  & \textbf{0.28} & \textbf{0.90} & 0.37                           & 1.13                           & \textbf{0.84} & 1.16                           & 2.28                           & 2.12                           \\ \hline
\multicolumn{1}{|l|}{Predictive - 200 sets} & 0.29                           & \textbf{0.90} & 0.36                           & 1.11                           & 0.91                           & \textbf{1.15} & 2.14                           & 1.97                           \\ \hline
\end{tabular}
\end{table*}

While optimizing the objective function, it is also important to ensure that the gap in admission rates is decreased compared to when we do not intervene.
Note that no intervention (i.e., no bonus, $b = 0$) is almost guaranteed to lead to the lowest utility loss, as utility is typically largest when no affirmative action policy is applied.
Table~\ref{tab:baselines_difference_intervention_absolute_spd} compares the difference in the admission rate gaps, measured as the absolute value of \gls{SPD} with and without a bonus policy $b$: $|\text{SPD}_b| - |\text{SPD}_0|$.
Negative values thus indicate a lower admission rate gap through the intervention -- which is desirable.
In general, we can see that the values suggested through more application sets again exhibit less variance.

\begin{table*}[h]
\scriptsize
\centering
\caption{Difference in absolute \gls{SPD} relative to no intervention. Lower values are better.}
\label{tab:baselines_difference_intervention_absolute_spd}
\begin{tabular}{l|l|l|l|l|l|l|l|l|}
\cline{2-9}
                                            & \multicolumn{4}{l|}{All programs}                                                                                                           & \multicolumn{4}{l|}{Consistently unequal}                                                                                                   \\ \cline{2-9} 
                                            & \multicolumn{2}{l|}{Gender}                                          & \multicolumn{2}{l|}{Income}                                          & \multicolumn{2}{l|}{Gender}                                          & \multicolumn{2}{l|}{Income}                                          \\ \hline
\multicolumn{1}{|l|}{Strategy}              & Mean                              & SD                               & Mean                              & SD                               & Mean                              & SD                               & Mean                              & SD                               \\ \hline
\multicolumn{1}{|l|}{Historical - 1 year}   & 0.0014                            & 0.0339                           & -0.0013                           & 0.0383                           & 0.0028                            & 0.0922                           & -0.0774                           & 0.1000                           \\ \hline
\multicolumn{1}{|l|}{Historical - 3 years}  & \textbf{-0.0005} & 0.0269                           & -0.0034                           & 0.0266                           & -0.0153                           & 0.1069                           & 0.0983                            & \textbf{0.0704} \\ \hline
\multicolumn{1}{|l|}{Historical - 5 years}  & 0.0005                            & 0.0234                           & \textbf{-0.0037} & 0.0270                           & -0.0037                           & 0.0858                           & \textbf{-0.0849} & 0.0827                           \\ \hline
\multicolumn{1}{|l|}{Predictive - 50 sets}  & -0.0003                           & \textbf{0.0124} & -0.0020                           & \textbf{0.0210} & \textbf{-0.0156} & 0.0586                           & -0.0625                           & 0.0822                           \\ \hline
\multicolumn{1}{|l|}{Predictive - 200 sets} & -0.0002                           & 0.0126                           & -0.0023                           & 0.0215                           & -0.0135                           & \textbf{0.0575} & -0.0670                           & 0.0817                           \\ \hline
\end{tabular}
\end{table*}

The reason for this lies in the nature of the predictive approach which is more conservative in its suggestions.
To see this, we compare the number of bonus points given to each program under the different strategies (see Table~\ref{tab:baselines_bonus_values}).
What is evident is that the more application sets a strategy bases its suggestions on, the smaller the proposed bonus values become and the less variance they show across all programs.
The bonus values suggested by the predictive approaches are thus closest to 0 and vary the least.
The optimal policy in general gives more bonus points.
Even though the strategy based on last year's historical data has similar statistics, the previous analysis has shown that this strategy often performs worse than the other strategies.
This underlines the need for a more conservative design strategy.

\begin{table}[h]
\scriptsize
\centering
\caption{Comparison of the number of bonus points given by different strategies.}
\label{tab:baselines_bonus_values}
\begin{tabular}{l|l|l|l|l|}
\cline{2-5}
                                            & \multicolumn{2}{l|}{Gender} & \multicolumn{2}{l|}{Income} \\ \hline
\multicolumn{1}{|l|}{Strategy}              & Mean         & SD           & Mean         & SD           \\ \hline
\multicolumn{1}{|l|}{Historical - 1 year}   & 2.31         & 6.33         & 2.49         & 6.49         \\ \hline
\multicolumn{1}{|l|}{Historical - 3 years}  & 1.85         & 4.06         & 1.90         & 4.58         \\ \hline
\multicolumn{1}{|l|}{Historical - 5 years}  & 1.50         & 3.25         & 1.77         & 4.28         \\ \hline
\multicolumn{1}{|l|}{Predictive - 50 sets}  & 0.97         & 2.70         & 1.25         & 3.96         \\ \hline
\multicolumn{1}{|l|}{Predictive - 200 sets} & 0.95         & 2.66         & 1.24         & 3.94         \\ \hline
\multicolumn{1}{|l|}{Ideal}                 & 1.76         & 6.24         & 2.21         & 6.78         \\ \hline
\end{tabular}
\end{table}

\spara{Programs with consistent inequalities}.
We observe in Table~\ref{tab:baselines_difference_intervention_absolute_spd} that some design strategies have averages above 0 and that the \gls{SD}s are large compared to the means.
Therefore, sometimes the strategies \emph{increase} the gap of admission rates compared to having no policy -- largely due to the unpredictability of admission rates (see Section~\ref{sec:setting}).

In practice, affirmative action policies should not be deployed for all programs, but only sparingly for programs that have consistently unequal admission rates.
Hence, we consider programs that (i) have unequal admission rates in each of the three most recent years, where (ii) the admission rate is always lower for the same subgroup, and (iii) the differences are above 0.1, i.e., strong differences, for at least two out of the three years.
This results in 12 programs to which a gender policy is applicable and 29 programs for income policies.
This means that income tends to generate more consistent and stronger admission rate disparities than gender.
Anecdotally, we note that programs with consistent inequalities tend to be large programs with many high-scoring applicants, such as medicine or engineering.

Table~\ref{tab:baselines_differences_objective_function} and \ref{tab:baselines_difference_intervention_absolute_spd} include the values for the filtered programs in the columns on the right.
The findings are similar to what we had previously observed for all programs.
The predictive policy suggestions again exhibit lower variance with the exception of the income policies.

In general, we note that strategies based on predictive models are less likely to harm an underrepresented group in cases where we observe ``flips'' in the underrepresentation that sometimes happen from one year to the next one.
These changes can to some extent, but not entirely, be anticipated and addressed with more conservative policy suggestions.

%% file: chapters/05_conclusions.tex
\section{Conclusions}\label{sec:discussion}

We compared a predictive approach to simpler design strategies based on averaging retrospectively optimal bonuses over 1-5 years of historical data.
The latter are more likely to over-correct the differences in admission rates.
Our proposed predictive approach mostly avoids this pitfall through more conservative suggestions.
In practice, while the predictive strategy tends to be more robust than the simpler approaches, a simpler approach based on sufficient historical data (e.g., the last five years) might in practice be preferable.
Predictive methods are, however, advantageous if historical data is only available for a few years or if only aggregate statistics are accessible.

An aspect that is left for future research is the effect of the announcement of affirmative action policies on the application behavior.
Existing research (see, e.g.,~\cite{women-competition, dickson2006does, change-in-application-behavior}) suggests that the mere existence of affirmative action policies might encourage students from disadvantaged groups to apply.
It is therefore imaginable that less bonus points may suffice to achieve the desired effect: one more reason for preferring conservative strategies.